# Equity Implications of Net-Zero Emissions: A Multi-Model Analysis of Energy Expenditures Across Income Classes Under Economy-Wide Deep Decarbonization Policies

John Bistline[a,1], Chikara Onda[b], Morgan Browning[c], Johannes Emmerling[d], Gokul Iyer[e], Megan Mahajan[f], Jim McFarland[c], Haewon McJeon[g], Robbie Orvis[f], Francisco Ralston Fonseca[a], Christopher Roney[a], Noah Sandoval[h], Luis Sarmiento[d], John Weyant[i], Jared Woollacott[j], Mei Yuan[k]

[a] Electric Power Research Institute, 3420 Hillview Avenue, Palo Alto, California 94304, USA
[b] U.S. Department of Energy, 1000 Independence Avenue SW, Washington, DC 20585, USA
[c] U.S. Environmental Protection Agency, 1200 Pennsylvania Avenue NW, Washington, DC 20004, USA
[d] RFF-CMCC European Institute on Economics and the Environment, Centro Euro-Mediterraneo sui Cambiamenti Climatici, Via Bergognone 34, 20144 Milan, Italy
[e] Pacific Northwest National Laboratory, 5825 University Research Court, College Park, Maryland 20740, USA
[f] Energy Innovation, 98 Battery Street #202, San Francisco, California 94111, USA
[g] KAIST Graduate School of Green Growth & Sustainability, Daejeon, 34141, Republic of Korea
[h] National Renewable Energy Laboratory, 15013 Denver W Parkway, Golden, Colorado 80401, USA
[i] Stanford University, 475 Via Ortega, Stanford, California 94305, USA
[j] RTI, 3040 E Cornwallis Road, Durham, North Carolina 27709, USA
[k] MIT Joint Program on the Science and Policy of Global Change, 77 Massachusetts Avenue, E19-411, Cambridge, Massachusetts 02139, USA

**Abstract**

With companies, states, and countries targeting net-zero emissions around midcentury, there are questions about how these targets alter household welfare and finances, including distributional effects across income groups. This paper examines the distributional dimensions of technology transitions and net-zero policies with a focus on welfare impacts across household incomes. The analysis uses a model intercomparison with a range of energy-economy models using harmonized policy scenarios reaching economy-wide, net-zero $CO_2$ emissions across the United States in 2050. We employ a novel linking approach that connects output from detailed energy system models with survey microdata on energy expenditures across income classes to provide distributional analysis of net-zero policies. Although there are differences in model structure and input assumptions, we find broad agreement in qualitative trends in policy incidence and energy burdens across income groups. Models generally agree that direct energy expenditures for many households will likely decline over time with reference and net-zero policies. However, there is variation in the extent of changes relative to current levels, energy burdens relative to reference levels, and electricity expenditures. Policy design, primarily how climate policy revenues are used, has first-order impacts on distributional outcomes. Net-zero policy costs, in both absolute and relative terms, are unevenly distributed across households, and relative increases in energy expenditures are higher for lowest-income households. However, we also find that recycled revenues from climate policies have countervailing effects when rebated on a per-capita basis, offsetting higher energy burdens and potentially even leading to net progressive outcomes. Model results also show carbon Laffer curves, where revenues from net-zero policies increase but then decline with higher stringencies, which can diminish the progressive effects of climate policies. We also illustrate how using annual income deciles for distributional analysis instead of expenditure deciles can overstate the progressivity of emissions policies by overweighting revenue impacts on the lowest-income deciles.

---

[1] Electric Power Research Institute, 3420 Hillview Avenue, Palo Alto, CA 94304. Corresponding author. Email: jbistline@epri.com. Phone: 650-855-8517.

# 1. Introduction

As companies, states, and countries target net-zero emissions around midcentury, there are questions about how these goals alter household welfare. Decarbonization policies, including those to reach net-zero emissions across the economy, have different impacts on households depending on characteristics such as income and energy use. Policy design influences changes in the price of carbon-intensive goods and consumption behavior, which in turn affects household welfare and emissions.

There is a large and growing literature on environmental justice and energy equity (e.g., recent studies such as [1, 2, 3, 4, 5] and literature reviews such as [6, 7, 8]). The literature on the distributional effects of climate policy and broader environmental policy dates back several decades [9, 10]. There are many single-model studies on equity outcomes under climate policies [11, 12], though most do not reach net-zero emissions levels. Multi-model studies of distributional impacts of climate policy similarly do not reach net-zero emissions and often focus on general equilibrium models with less detailed energy systems modeling [13]. As articles examining equity issues and net-zero emissions are often qualitative [14], we contribute to this literature by providing the first multi-model study of the distributional impacts of net-zero emission policies, focusing on changes in direct energy expenditures across income classes. Multi-model studies are valuable for testing the robustness of insights across different input assumptions and model frameworks.

This paper examines the distributional dimensions of technology transitions and net-zero policies with a focus on welfare impacts across household incomes. The analysis uses a model intercomparison exercise with a range of energy-economy models using harmonized policy scenarios reaching economy-wide, net-zero $CO_2$ emissions across the United States in 2050. Some models report income-related variables directly, but to facilitate contributions from a greater variety of models, we also pair outputs from energy-economic models with survey data on energy expenditures. This cross-cutting paper is part of a larger Energy Modeling Forum (EMF) 37 study investigating deep decarbonization and high electrification in North America [15]. The study aims to identify differences in model structures and input assumptions that explain differences in outputs across models. We also investigate the effects of policy implementation on energy expenditures across income groups, showing changes with and without potential revenue recycling, as well as comparing a standards-focused policy portfolio to reach net-zero $CO_2$ targets with carbon pricing.

We focus on outcomes across different households by income, including policy changes in direct energy expenditures and energy burdens. Although there are many additional equity and incidence questions around the costs and benefits of net-zero policies—health impacts,[2] energy insecurity, employment, asset ownership, risks to local economies, and others—effects across income groups are a focal issue of affordability for policymakers and other stakeholders. In addition, many of these additional equity and incidence questions are directly related to income, since it can act as a proxy for other characteristics that could be impacted by distributional injustices such as health and race [16, 17]. Excess energy burdens are important given their impacts on physical and mental health, comfort, education, job performance, and community development [18]. Households with high energy burdens are more likely to be caught in poverty cycles [18], and households may forgo needed energy use to reduce energy bills [19]. For instance, one in five U.S. households reports reducing or forgoing necessities such as food or medicine to pay energy bills [20]. By reporting on the distributional impacts of reaching economy-wide, net-zero $CO_2$ emissions for households at different income levels, we can inform discussions about the wider impact of these scenarios on not only the economic but also the physical, mental, and emotional welfare of the most vulnerable.

---
[2] Note that a complementary EMF 37 study focuses on air pollutant emissions.

## 2. Methods

### 2.1. Approach

The analysis uses a two-track approach. In Section 2.2, we first pair the outputs from energy system models with household expenditure data from the Consumer Expenditure Survey (CEX) conducted by the U.S. Bureau of Labor Statistics. Our analysis modifies the approach of Cullenward, et al. (2016) [21] by taking aggregate EMF 37 model projections across each residential fuel and consumption category and then scaling these expenditures by CEX expenditure data:

$$\epsilon_{fi}^t = \left(\frac{\epsilon_f^t}{\epsilon_f^0}\right) \cdot \epsilon_{fi}^0 = \left(\frac{p_f^t q_f^t}{p_f^0 q_f^0}\right) \cdot \epsilon_{fi}^0 \quad (1)$$

Where $\epsilon_{fi}^t$ is expenditure in period $t$ for fuel $f$ and household type $i$ (i.e., income classes), $p_f^t$ is the price of $f$ in period $t$, and $q_f^t$ is the quantity of $f$ consumed in $t$ ($0$ designates values in the base year). The $(\epsilon_f^t/\epsilon_f^0)$ term is the model-specific fuel expenditure (normalized by base year spending), and $\epsilon_{fi}^0$ is the base year expenditure by income class from the CEX. This approach incorporates model-specific changes over time in prices and quantities of household expenditures, which can shift from technological change (e.g., transport electrification) and policy (e.g., carbon pricing increasing fuel prices). This means that the change in total direct energy expenditure (i.e., direct component of total impacts) would be:

$$\Delta\epsilon_{fi}^t = \epsilon_{fi}^t\Big|_{pol} - \epsilon_{fi}^t\Big|_{ref} \quad (2)$$

Where $\epsilon_{fi}^t\Big|_s$ is the expenditure in period $t$ for fuel $f$ evaluated in scenario $s$ ("pol" is the policy scenario, and "ref" is the reference without the policy). Note that changes in direct fuel consumption are only a subset of total costs and do not account for changes in expenditures on other goods and services.

CEX data report household expenditures—including energy—across income levels, geographies, and other demographic variables. We use CEX Public Use Microdata from 2019 and aggregate across consumer units (i.e., household-level statistics in the CEX microdata) by income decile for many figures. Figure 1 shows how energy spending as a share of pre-tax income is highest for low-income households, though absolute spending increases in income.[3] Electricity spending is relatively flat across income classes ($1,400 to $1,600), except for the lowest and highest income households. Petroleum is the largest energy-related expense for many households and is nearly three times higher between the lowest- and highest-income households.

---

[3] The "Other Transport" category includes expenditures on public transport, rental cars, and airfare.

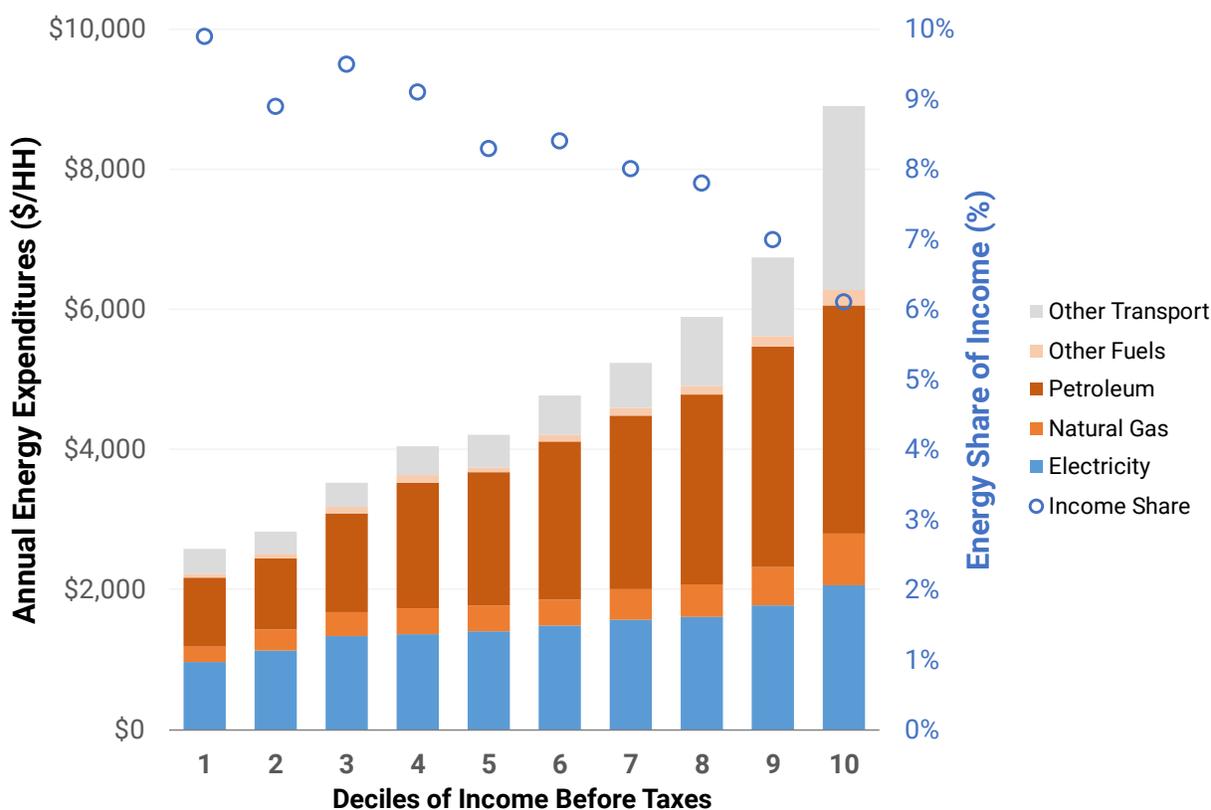

**Figure 1. Annual household energy expenditure by fuel across income classes (in 2019 U.S. dollars).** Data come from the U.S. Consumer Expenditure Survey 2019 cross-tabulated data.

The second approach is to compare the output of models that explicitly represent income groups, summarized in Section 3.3. However, since the number of participating models with these structural classes is limited (Section 2.3), the analysis focuses on the approach of pairing model output with survey microdata that are disaggregated by income.

*2.2. Scenarios*

This analysis focuses on two policy scenarios:

- **Reference:** The counterfactual reference scenario assumes that there are no new climate or energy policies after early 2022. This scenario includes on-the-books state and federal policies and incentives but excludes the Inflation Reduction Act. As described in [15], the coverage of existing policies varies by model and depends on regional disaggregation, especially since some models represent the U.S. as a single region (Table 1). Reference or baseline scenarios are useful in policy analysis to isolate the potential impacts of proposed policies on costs and emissions outcomes with all other factors held constant.[4]

---

[4] Uncertainty about future policies, markets, and technologies mean that an analysis may adopt multiple reference scenarios and then investigate how such alternate baselines can influence policy analysis [32, 33, 34]. However, in all instances, policy evaluation requires the specification of a counterfactual reference to attribute impacts to policies rather than other drivers that may be simultaneously changing over time.

- **0by50:** This scenario reaches net-zero national $CO_2$ emissions by 2050 and requires a linear $CO_2$ reduction between 2020 and 2050. The target is defined in terms of $CO_2$ emissions only, including technological carbon dioxide removal (CDR) and natural carbon management.[5] Most models represent these emissions constraints through a cap on net $CO_2$ emissions, where revenues are recycled to consumers with a lump-sum distribution on a per-capita basis.[6] This harmonized revenue recycling assumption helps to maintain consistency across models. This emissions cap is layered on top of on-the-books federal and state policies, incentives, and standards. A notable exception in implementation is the Energy Policy Simulator (EPS), which uses a combination of policies to reach the net-zero targets, including standards, instead of a $CO_2$ constraint. The contrast of policy implementation in the study illustrates potential effects on household energy expenditures and reflects uncertainty about how such deep decarbonization goals can be reached and the political economy of different instruments.[7]

For more information on the scenario design and participating models, refer to the EMF 37 overview paper [15]. Although policy assumptions are harmonized across models, other assumptions—including fuel prices (Figure 15)—are unharmonized. Scenarios should not be interpreted as reflections of the likelihoods of potential outcomes or of the technology, market, or policy preferences of the modelers.

We also conduct several side analyses:

- Results are generally shown with and without recycled $CO_2$ revenues, given how there is uncertainty about how these revenues could be used.
- We show distributional impacts primarily across annual income deciles, which is commonly used in policy analysis because of the prevalence of these data. However, since the literature suggests that annual expenditures and consumption may be a better proxy for lifetime income and welfare impacts, we conduct a sensitivity where expenditure declines are used instead of annual income.
- The results in the paper generally use model projections for policy changes in prices and quantities, per Equation 1. Given the uncertainty of price-induced responses across different income classes, especially for lowest-income households, we investigate an extreme case where energy consumption is perfectly inelastic for the lowest-income households, with changes in expenditures scaled only by prices.

---

[5] Note that the U.S. Nationally Determined Contribution is to reach net-zero greenhouse gas (GHG) emissions by 2050. The EMF 37 study includes a net-zero GHG sensitivity, but we omit this scenario from the analysis given the limited number of models submitting this sensitivity.
[6] Payments for CDR are assumed to be subtracted from carbon pricing revenues (i.e., net emissions revenues are distributed to households). However, the land sink emissions from the reference scenario are assumed not to require payments in the 0by50 scenario, which are approximately -800 Mt-$CO_2$/yr in many models (except for US-REGEN, which has a carbon land sink of about -200 Mt-$CO_2$/yr). These assumptions imply that revenues can be non-zero when net-zero $CO_2$ is reached. Revenues are distributed on a per-capita basis based on the average number of occupants, which varies by income.
[7] States and other jurisdictions in the U.S. that have made net-zero emissions pledges largely have not yet passed the policies, regulations, and other drivers necessary to reach these targets, so there is considerable room for debate about which future policy portfolios might be used. Although about 40% of greenhouse gas emissions in the OECD are currently subject to carbon pricing [31], the public opinion literature suggests variation in and context-specific support for carbon pricing vis-à-vis alternate policy approaches [30].

*2.3. Models*

The models in this analysis are a subset of EMF 37 models that reported variables related to residential prices and quantities. Participating models span a wide range of scopes, approaches, and structures (Table 1). One model (WITCH) reports income-related variables explicitly.

**Table 1.**
**Participating models and key features.** Coverage and equilibrium approach: PE, partial equilibrium; LP, linear program; IAM, integrated assessment model.

| Analysis Abbreviation | Model(s) | Analysis Institution | Coverage and Approach | Income Classes? | Geographic Coverage | Link |
|---|---|---|---|---|---|---|
| **ADAGE** | Applied Dynamic Analysis of the Global Economy | RTI International | Economy-wide, computable general equilibrium | No | Global with 8 regions | Link |
| **EPS** | Energy Policy Simulator (EPS) | Energy Innovation | Economy: System dynamics; logit choice | No | Single region[8] | Link |
| **GCAM** | Global Change Analysis Model | PNNL | Economy: Logit choice | No | Single region entire U.S. | Link |
| **GCAM-USA** | Global Change Analysis Model for USA | UMD-CGS | Economy: Logit choice | No | 50 U.S. states and D.C. | Link |
| **US-REGEN** | Regional Economy, Greenhouse Gas, and Energy | EPRI | Energy end use: Lagged logit choice; Power: Least-cost LP | No | 16 U.S. regions | Link |
| **USREP-ReEDS** | U.S. Regional Energy Policy, Regional Electricity Deployment System | MIT, NREL, RTI International | Economy-wide computable general equilibrium (USREP) with linkage to partial equilibrium (ReEDS) | Yes | 12 U.S. regions | Link1 Link2 |
| **WITCH** | World Induced Technical Change Hybrid | RFF-CMCC EIEE | General equilibrium hybrid top-down/bottom-up IAM | Yes | Global with 17 regions | Link |

Although models represent energy systems in detail, they differ in many respects, such as their ability to explicitly look at income, general equilibrium effects, and degree of foresight.

*2.4. Caveats*

---
[8] Separate state models for the contiguous U.S. are available for EPS.

There are several caveats to keep in mind when interpreting the analysis results:

- The analysis focuses on changes in direct fuel consumption, which is a subset of total policy costs. The literature suggests that indirect costs (e.g., policy-induced increases in prices of non-energy goods and services) may be comparable to direct costs [22]. The focus on fuel costs also omits changes in capital and maintenance costs of end-use technologies, which differ between electric and fossil options. These omissions mean that total costs will be higher than the direct fuel costs shown in our study.
- Although there is considerable variation within income classes, including heterogeneity of adaptive behavior (e.g., substitution, energy efficiency improvements) by income class, we focus on "vertical equity" in our analysis.[9] We illustrate possible "horizontal equity" effects in Section 4, which suggest that individual household impacts may be higher or lower than the average within a decile.
- Per Section 2.1 (Equation 1), many results in this study scale model-specific changes in fuel prices and demand by survey microdata for expenditures across different income deciles. However, there may be variation in responses across and within income classes, especially for lower-income households (e.g., limited disposable income or financing constraints for lower-income households may prevent from purchasing higher-efficiency options or fuel switching, even with lower lifetime costs of ownership). These considerations motivate the side case at the end of Section 3.2.
- Revenue recycling details of future climate policies are highly uncertain. This analysis assumes that revenues are recycled to consumers with a lump-sum distribution on a per-capita basis. However, we show impacts across income classes with and without recycled revenues to provide bookend estimates for the effects of these revenues on households.
- The analysis assumes that income distributions do not change during the projection period.
- Welfare impacts do not include policy benefits related to mitigating climate change externality and associated co-benefits such as enhanced air quality, which are the focus of the EMF 37 ancillary impacts paper.
- The results are contingent on assumptions about technologies, markets, and policies. Given uncertainties about projections for these values, results should not be taken as predictive, especially for longer-run outputs; however, comparative insights can be useful by looking at relative changes across scenarios.

## 3. Results

This section begins with aggregate national impacts across all models to provide insights into how household consumption and fuel expenditures change over time in these scenarios (Section 3.1). Next, we combine these outputs with CEX data to examine distributional impacts across income classes (Section 3.2). Finally, we compare the results across models that explicitly characterize households by income (Section 3.3).

---

[9] For instance, wealthy households may be better positioned to avoid energy price increases by shedding discretionary demand or to take advantage of energy price declines.

*3.1. Aggregate National Welfare Impacts Across Models Over Time*

The results of the model indicate that the electricity demand per household is expected to increase over time (Figure 2). There is variation in the extent of these increases with generally higher demand under 0by50 relative to the reference (rising by 21-105% by 2050 from 2020 levels in the reference vs. 37-108% in the 0by50 scenario). Petroleum use similarly declines in part due to increases in transport electrification, especially under net-zero $CO_2$ policies.

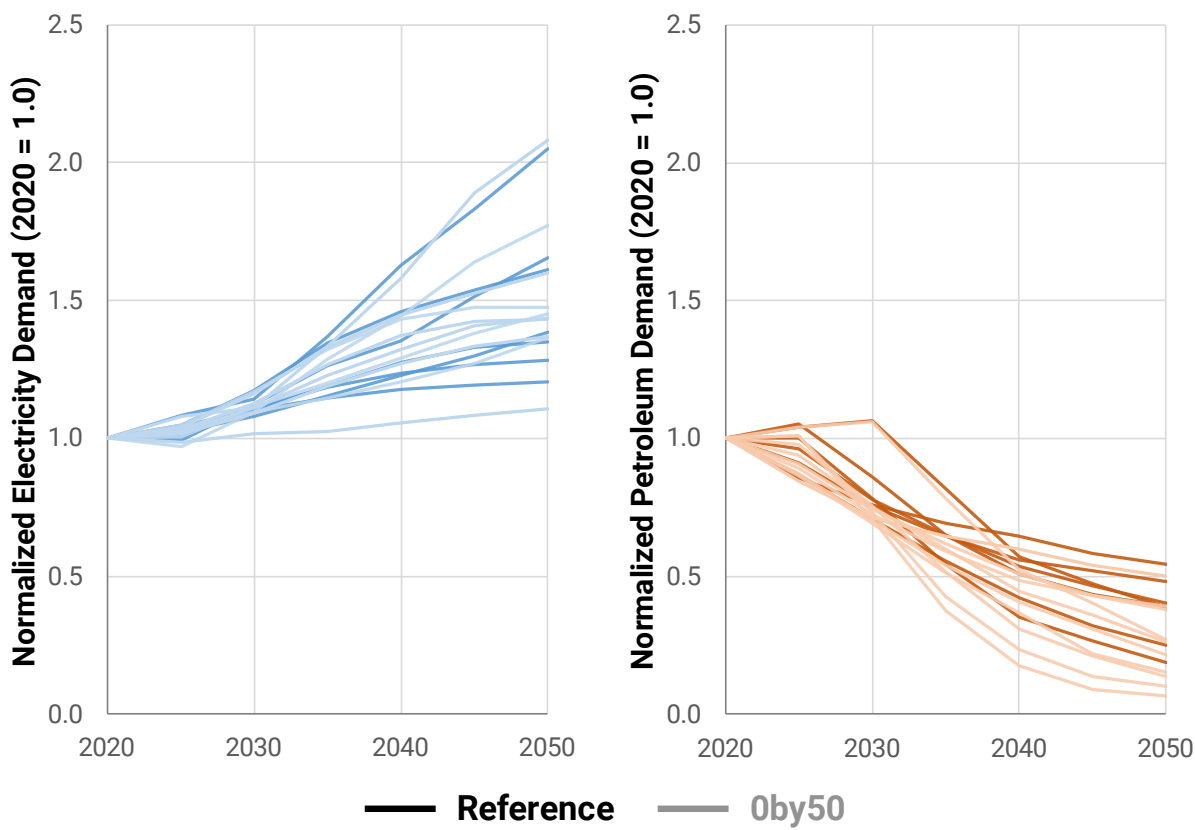

**Figure 2. Demand for residential electricity (left panel) and petroleum (right panel) over time.** The lines show individual model results for the reference scenario (dark lines) and 0by50 scenario (light lines). Values are normalized to their 2020 levels.

Combining changes in fuel consumption with residential prices, Figure 3 compares average household energy expenditures by fuel across models. Models generally agree that direct energy expenditures for households decrease between 2020 and 2050. That is, energy becomes more affordable over time in absolute terms. Relative declines would be even more prominent as incomes rise. Many models show reductions in total household energy service costs from electrification in the reference relative to current levels, even with higher electricity prices and spending. The reference shows a strong trend toward light-duty vehicle electrification over time (Figure 2). There is general agreement in most models that net-zero policies increase total energy expenditures relative to the reference. The lower household expenditures in EPS stem from the model implementation of the 0by50 scenario, which uses a combination of policies, including standards, instead of the approaches used in other models, such as $CO_2$ constraints or prices, which increase prices of emissions-intensive fuels. Higher expenditures in the other models reflect fossil fuel price increases (inclusive of carbon pricing, per the Appendix) outweighing decreases in petroleum and natural gas consumption from electrification and efficiency under the net-zero policy.

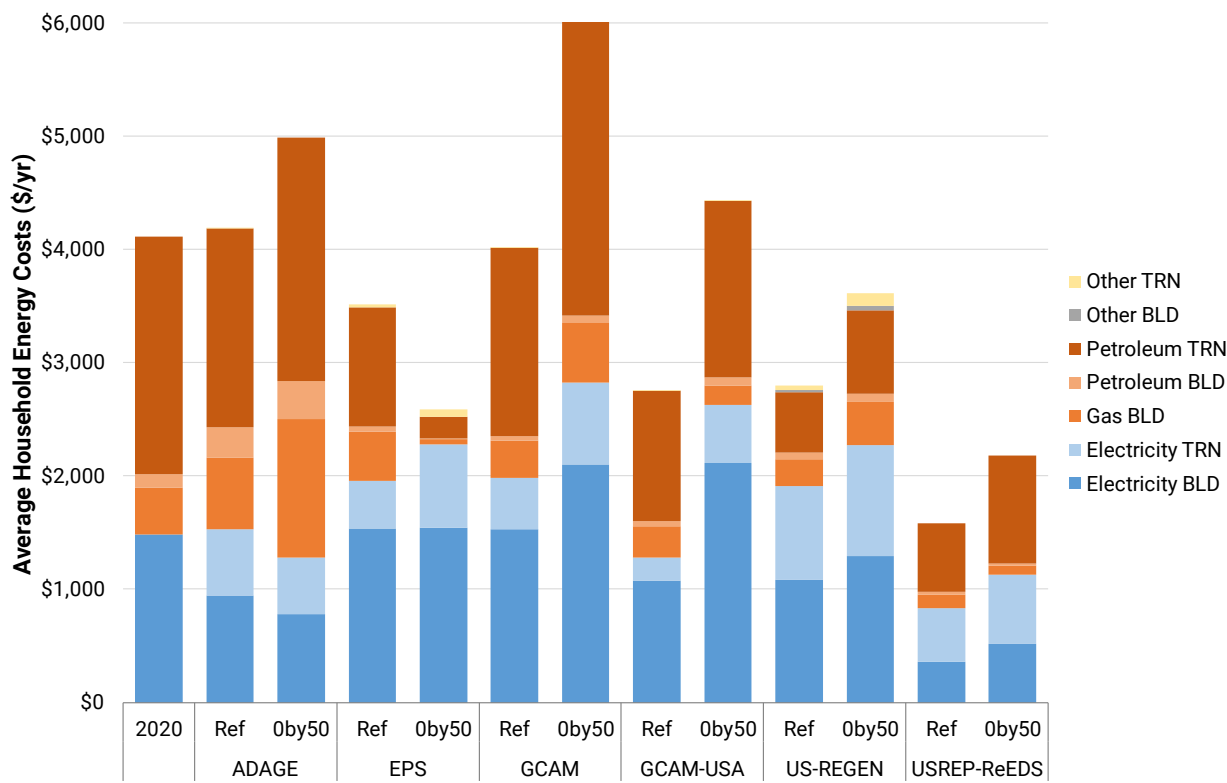

**Figure 3. Cross-model comparison of average household energy expenditures by category in 2050.** Bars show values under the reference scenario (Ref) and net-zero $CO_2$ by 2050 (0by50). Fuel expenditure is decomposed into transportation (TRN) and buildings (BLD) expenditures.

Carbon prices are shown over time in Figure 4, which are model outputs that reflect the cost of the last unit of $CO_2$ reduced to meet the emissions cap constraint. As discussed in the EMF 37 overview paper [15], the range of $CO_2$ shadow prices over time reflects a combination of factors,[10] including:

- Model structure (e.g., spatial and temporal resolution)
- Coverage of technology options and fuel transformation pathways (e.g., technological carbon removal, low-carbon energy carriers)
- Assumptions about exogenous and endogenous behavioral changes (e.g., price-responsive energy efficiency, service demand elasticities)
- Input assumptions about technological cost and performance (e.g., capital costs of supply- and demand-side options over time, feedstock supply curves such as biomass, efficiency assumptions for technologies)

---

[10] Carbon prices may be slower or faster with alternate assumptions about technological progress, complementary policies and regulations, or policy stringency.

The highest carbon prices in 2050 are from GCAM-USA, USREP-ReEDS, and GCAM models ($400-700/t-$CO_2$). Net-zero scenarios assume lump-sum distributions of recycled revenue to households on a per-capita basis after accounting for CDR-related payments. ADAGE has higher prices in the early periods when $CO_2$ is higher (and hence higher revenues) but then declines as the cost of the marginal abatement technology, direct air capture, decreases due to scale economies and technological progress, as well as due to fossil fuel burning equipment being replaced by low-emission alternatives.

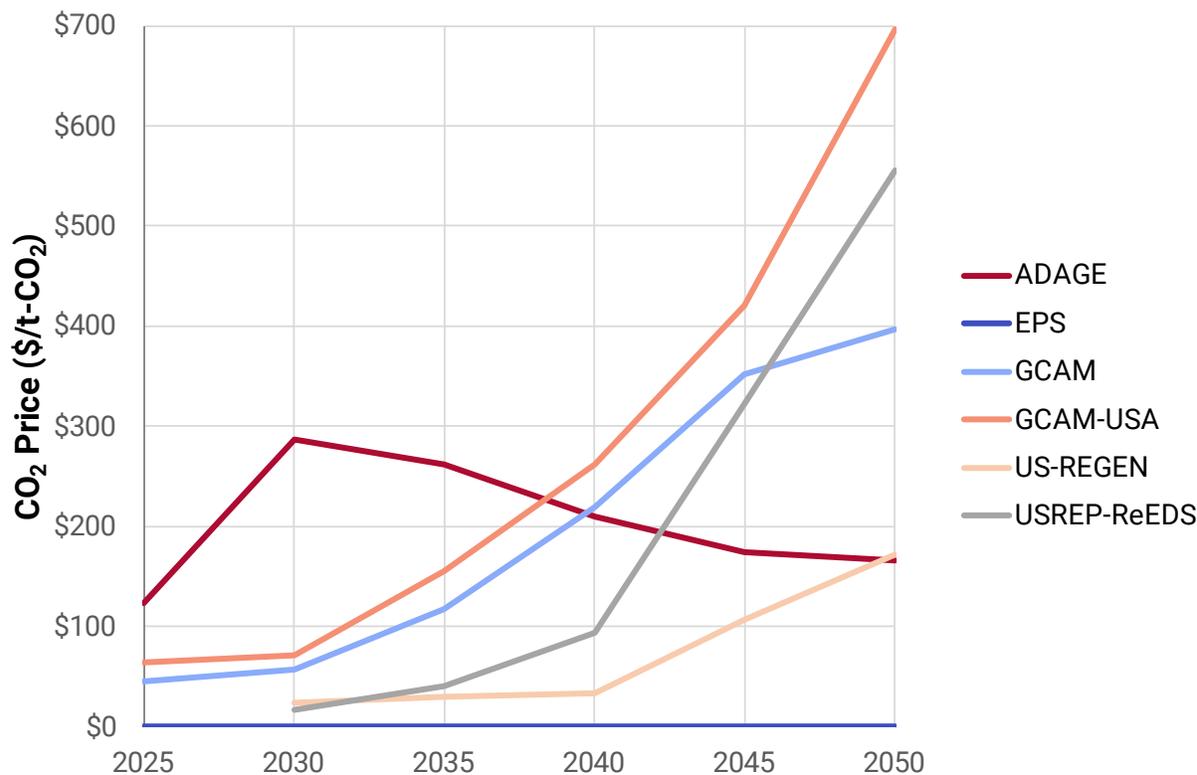

**Figure 4. $CO_2$ prices by model over time for the net-zero $CO_2$ by 2050 scenario.** These model outputs are the shadow prices on the emissions cap constraint over time. Note that EPS does not implement the net-zero target as a $CO_2$ constraint, which is the primary driver in other models.

Figure 5 illustrates how annual revenues from net-zero $CO_2$ policies exhibit carbon Laffer curves, where revenues initially increase in policy stringency and then decline with deeper decarbonization (i.e., higher tax rates). This non-linear relationship between tax revenue and decarbonization levels has been observed in several previous studies [23, 24]. This inverted U-shaped relationship is caused by lower $CO_2$ emissions and higher CDR deployment at higher policy stringencies, so although implicit $CO_2$ prices increase, these changes are more than offset by lower emissions and higher CDR outlays. Revenue paths vary by model and are positively correlated with changes in net household energy expenditures. Peaks of total revenue range from $110 billion per year to $990 B/yr,[11] and the decarbonization level where these maximum levels occur ranges from 46% to 74% reductions below their 2005 levels. These Laffer curve effects can diminish the progressive impacts of climate policies after revenue recycling at higher policy stringencies, where reducing emissions more than offset higher $CO_2$ prices. In other words, these dynamics lower the availability of funds to improve equity outcomes across income levels. Time trends for net policy revenues are shown in Figure 16.

---

[11] EPS does not report $CO_2$ revenues given how net-zero $CO_2$ policy is achieved through a combination of policies, including standards, rather than carbon pricing like the other models.

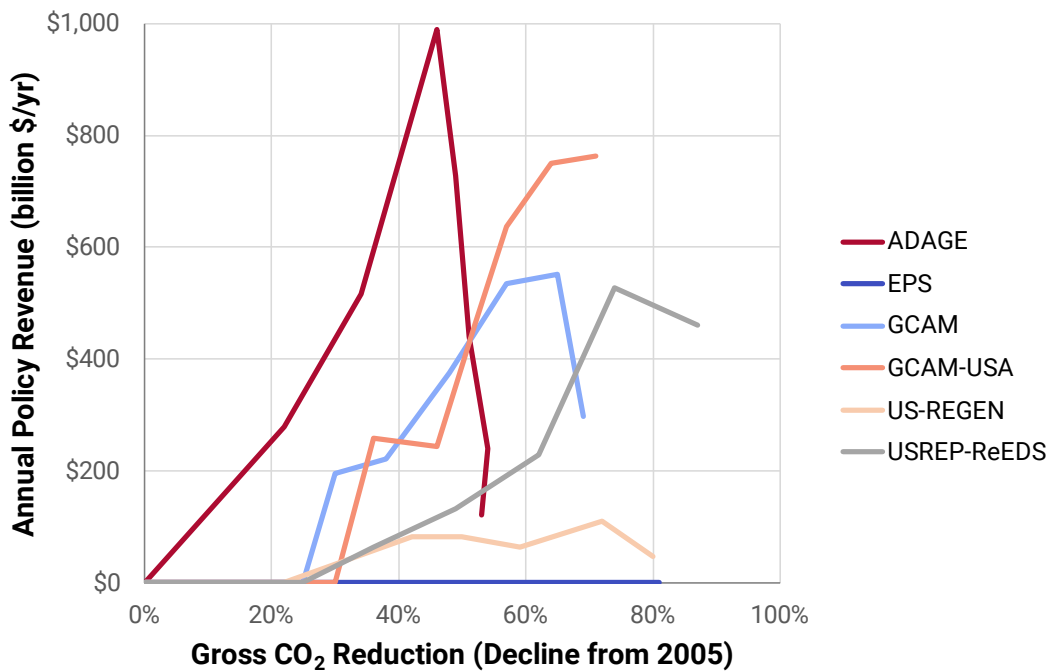

**Figure 5. Annual revenue from net-zero $CO_2$ policy by model.** Policy revenue is plotted against gross $CO_2$ reductions relative to 2005 levels.

*3.2. Distributional Impacts of Net-Zero Policies by Income Class*

The energy expenditures by income are shown in Figure 6. Models largely agree that electricity's share of expenditures increases over time, while total energy spending decreases, which is largely due to declining petroleum spending from electrification (foreshadowed in Figure 3). Effects of both vary by model. Models also differ in their overall magnitudes of energy expenditures.

There is considerable variation in the magnitudes of the recycled $CO_2$ revenues/rebates/dividends. These differences partially reflect differences in $CO_2$ price trajectories (Figure 4) and residual $CO_2$ emissions.[12] Rebating revenues can mitigate distributional consequences of the policy, and the size of these rebates increases for higher-income households, since they have more occupants.[13] These revenues are higher shares of income for poorer households. Models with low or zero revenues from carbon pricing (e.g., due to net-zero $CO_2$ being implemented with policies that are not wealth transfers) have larger net policy-induced impacts. The use of revenues matters as much as, if not more than, direct expenditure burdens and can vary depending on the policy design [25, 26, 27, 28].

---

[12] For instance, ADAGE has the highest rebates in 2030 due to its higher $CO_2$ prices, while GCAM-USA and USREP-ReEDS have the highest rebates and $CO_2$ prices in 2050.
[13] Scenarios assume a uniform distribution of policy revenues on per-capita basis.

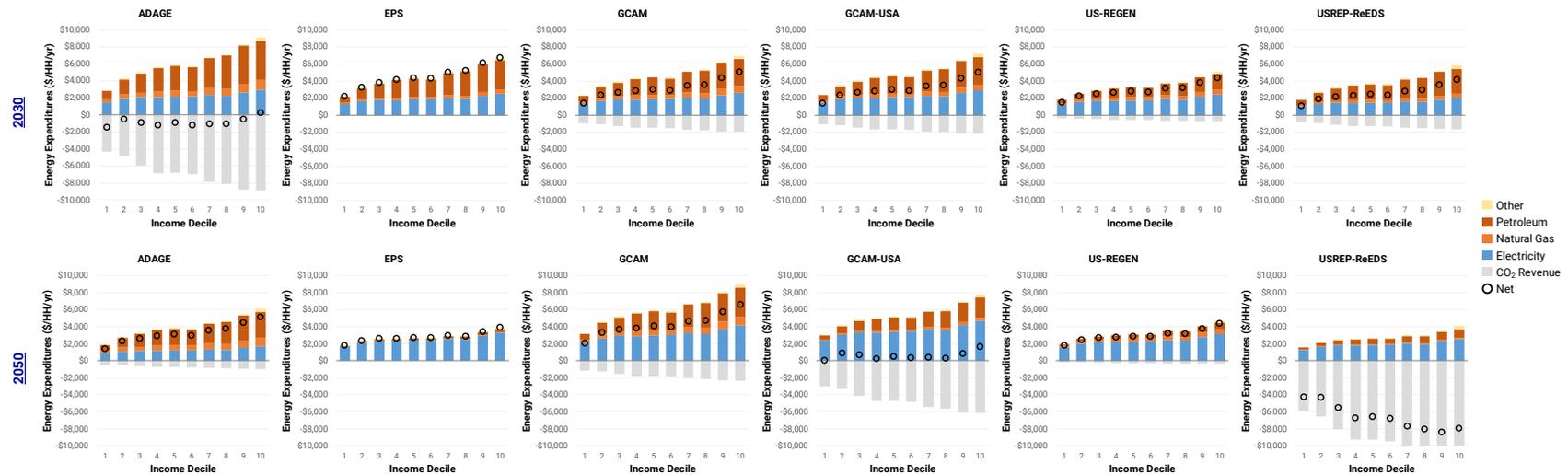

**Figure 6. Annual energy expenditures by fuel under the 0by50 policy scenario across income deciles.** Values are shown for 2030 (top row) and 2050 (bottom row) across models (columns).[14] Households are displayed by annual income deciles, sorted from lowest to highest income (1 through 10, respectively). $CO_2$ revenues are assumed to be recycled to consumers on a per-capita basis, as discussed in Section 2.2. EPS does not report $CO_2$ revenues given how net-zero $CO_2$ policy is achieved through a combination of policies, including standards, rather than carbon pricing like the other models.

---

[14] Note that the highest-income households in USREP-ReEDS in 2050 have $CO_2$ revenues of approximately $12,000 per household annually, but the bar is truncated to keep a common vertical axis on all panels.

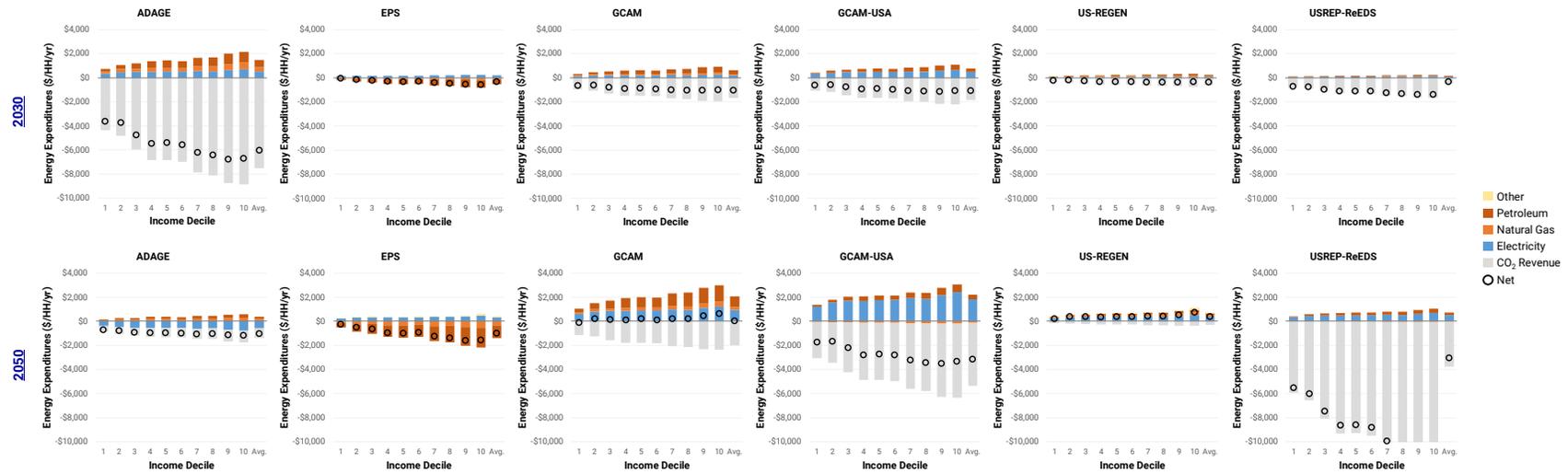

**Figure 7. Change in annual energy expenditures by fuel between the 0by50 and reference scenarios across income deciles (i.e., 0by50 costs minus reference costs).** Values are shown for 2030 (top row) and 2050 (bottom row) across models (columns).[15] Households are displayed by annual income deciles, sorted from lowest to highest income (1 through 10, respectively). $CO_2$ revenues are assumed to be recycled to consumers on a per-capita basis, as discussed in Section 2.2. EPS does not report $CO_2$ revenues given how net-zero $CO_2$ policy is achieved through a combination of policies, including standards, rather than carbon pricing like the other models.

---

[15] Note that the highest-income households in USREP-ReEDS in 2050 have $CO_2$ revenues of approximately $12,000 per household annually, but the bar is truncated to keep a common vertical axis on all panels.

Figure 7 shows the differences in annual energy expenditures between the reference and 0by50 scenarios. Increases with the net-zero policy scenario are above the horizontal axis and decreases below. With the net-zero policy, expenditures generally increase for electricity in many models and periods, reflecting both higher retail prices and policy-induced electrification. Models differ in whether net-zero policy increases or decreases expenditures on fossil fuels. These differences are mainly due to whether the net-zero target is implemented through carbon pricing. Five out of six models exhibit increases in spending on natural gas and petroleum (i.e., price increases from the carbon price outweigh declines in consumption under the net-zero policy). The rebate is sufficient to offset the increased direct energy expenditures for many models, years, and income classes.

Figure 8 shows changes in net energy expenditures as a share of the pre-tax household income, which is one measure of energy burden and impacts of net-zero policies on income inequality. Before accounting for recycled $CO_2$ revenues, net-zero policies are regressive[16] in five of six models and increase inequality between income classes. These disproportionate impacts on the lowest-income groups are due to the higher share of income that these households spend on energy (as shown in Figure 1). In other words, if net-zero policy does not facilitate compensating transfers to households, it may have a regressive impact.

However, this regressive incidence becomes progressive once the assumed per-capita rebates are taken into account, though the extent varies by model and year in Figure 8. Lump-sum transfers/rebates, such as a carbon dividend check or tax credit, can be progressive in that lower-income households see net gains from policy (i.e., their rebates exceed direct cost increases),[17] which are highest as a share of income. Carbon pricing can generate funds that can help reduce their distributional impacts (or achieve other policy goals), which is not necessarily the case with other policy designs such as revenue-neutral clean electricity standards [29]. However, the specifics of the policy design determine their relative incidence, since some carbon pricing policies do not raise revenues (e.g., cap-and-trade systems that freely allocate allowances), and non-carbon-pricing policies may reduce distributional impacts even if they do not raise revenue (e.g., vehicle tax credits with income eligibility limits). Note that these impacts can vary over time. Pre-transfer expenditure shares are larger for lower-income households and post-transfer smaller in 2030 across all models; however, the story is more mixed in 2050, as post-transfer shares are larger for lower-income households in two models.

---

[16] In incidence analysis, which is conducted to explore the distribution of benefits and/or costs across incomes classes, policies are regressive if costs as a share of income fall as income rises (and vice versa for progressive).
[17] Note that the magnitude of impact is less clear once indirect changes from the prices of non-energy goods and services are taken into account.

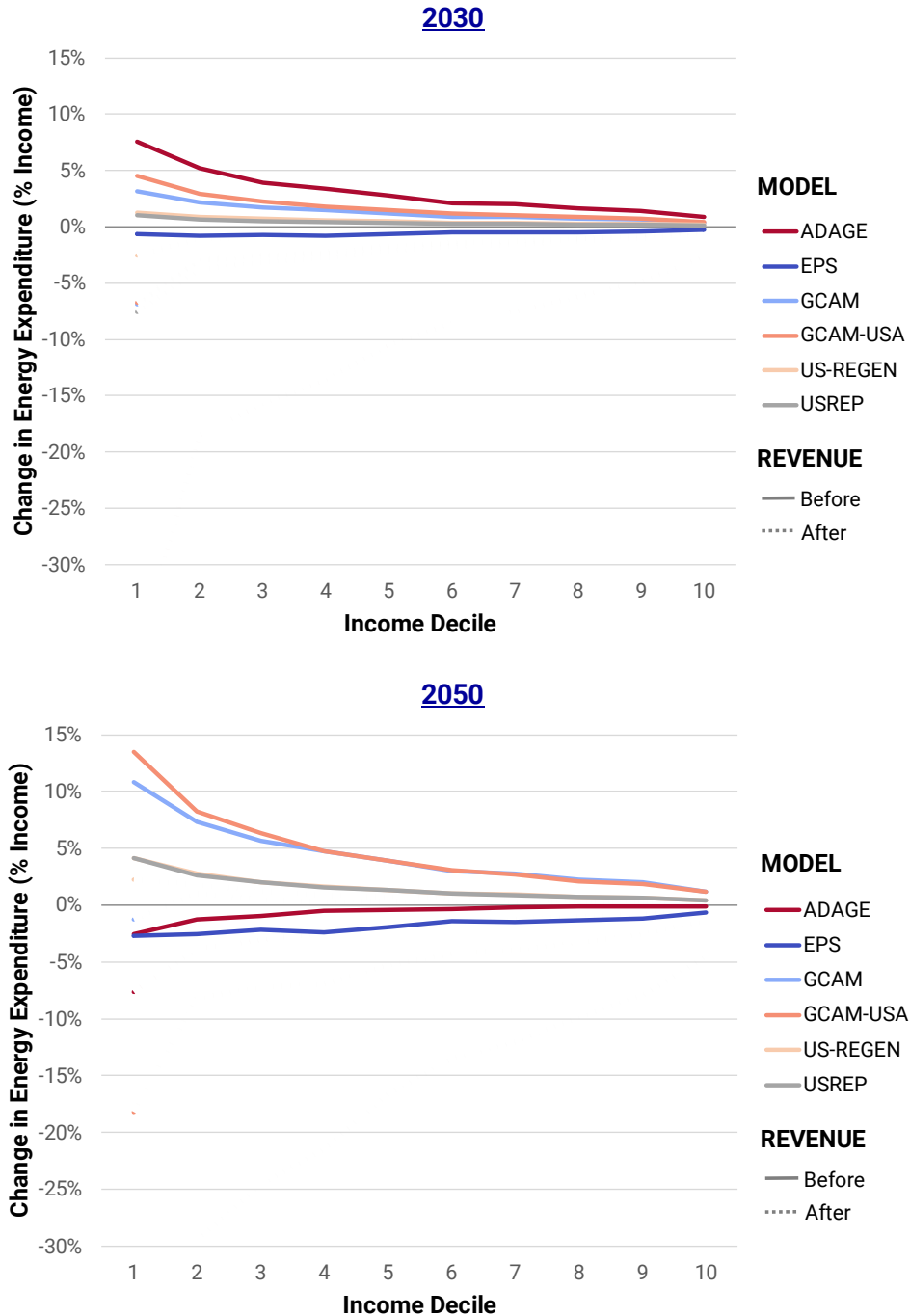

**Figure 8. Cross-model comparison of changes in annual energy expenditures between the 0by50 and reference scenarios across income deciles.** Values are shown as a share of pre-tax income before incorporating recycled $CO_2$ revenues (solid lines) and after these transfers (dotted lines). Panels show impacts in 2030 (top) and 2050 (bottom).

The previous results in this article show distributional impacts across deciles based on annual income. However, annual expenditures and consumption may be a better proxy for lifetime income and welfare impacts (i.e., retirees have higher consumption lifestyles than sorting by income suggests), though annual income is often used in policy analysis due to the prevalence of these data. To address this possible discrepancy, we use expenditure data from the same CEX dataset and the method discussed in Section 2.1. In other words, instead of scaling model projections by base year data across income classes, we scale results across expenditure classes.

Figure 9 compares 2050 impacts with expenditure classes against the earlier results in Figure 8 with income classes, where the vertical axis represents changes in energy expenditure as a share of annual income or overall expenditures. Using expenditure deciles flattens the distribution by controlling for households with low incomes but high expenditures (e.g., retirees living off savings). This comparison shows how using annual income deciles for distributional analysis instead of expenditures can overstate the progressivity of emissions policies by overestimating revenue impacts on the lowest-income deciles.

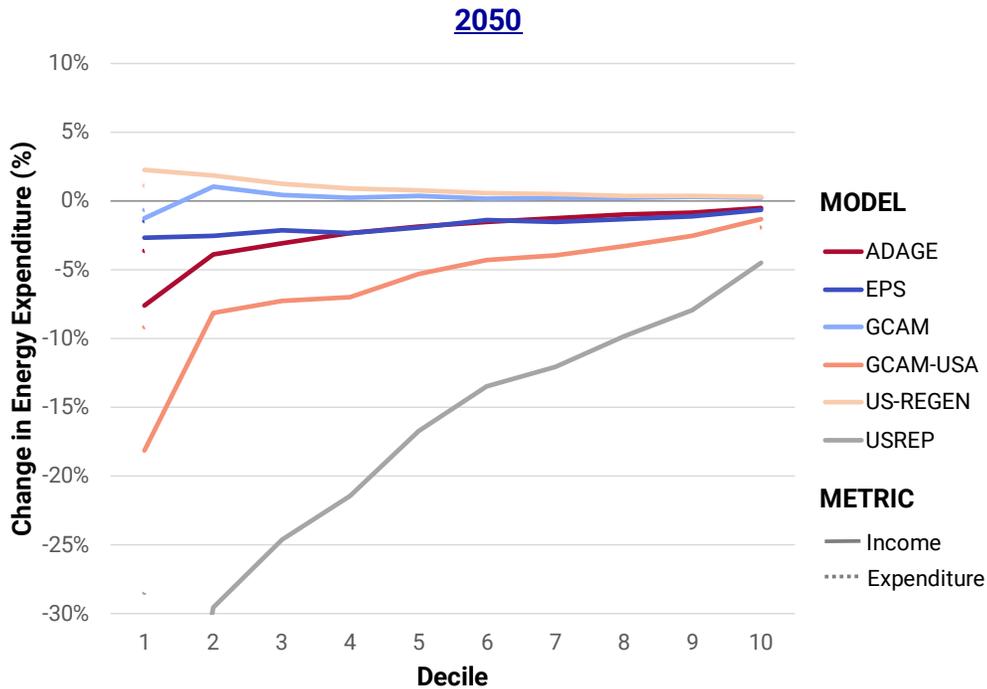

**Figure 9. Cross-model comparison of changes in annual energy expenditures between the 0by50 and reference scenarios across deciles in 2050.** Values are shown as a share of pre-tax income after incorporating $CO_2$ revenues for income deciles (solid lines) and expenditure deciles (dotted lines).

An additional sensitivity investigates the extreme case where energy consumption is perfectly inelastic for the lowest-income households, with changes in expenditures scaled only by prices:

$$\epsilon_{fi}^t = \left(\frac{p_f^t}{p_f^0}\right) \cdot \epsilon_{fi}^0 \qquad (3)$$

Where $\epsilon_{fi}^t$ is expenditure in period $t$ for fuel $f$ and household income $i$. On the one hand, scaling by quantity and price changes may underestimate the impact on expenditures to the extent that low-income households' energy expenditures are less elastic. On the other hand, scaling only by price overestimates spending impacts on low-income families by omitting price-induced margins of response.

Figure 10 shows the impacts of demand elasticities on energy expenditures for the lowest-income households. Assuming no behavioral response to climate policy leads to more significant fuel expenditures, though the extent and composition vary by model. The higher increases in expenditures with inelastic demand reflect the greater differences in fuel prices between the reference and 0by50 scenarios (Figure 15). Many models more than double their total energy expenditures with inelastic demand, and ADAGE is approximately three times higher due to its higher near-term $CO_2$ price (Figure 4), leading to more significant expenditure changes when demand for fuels is inelastic.

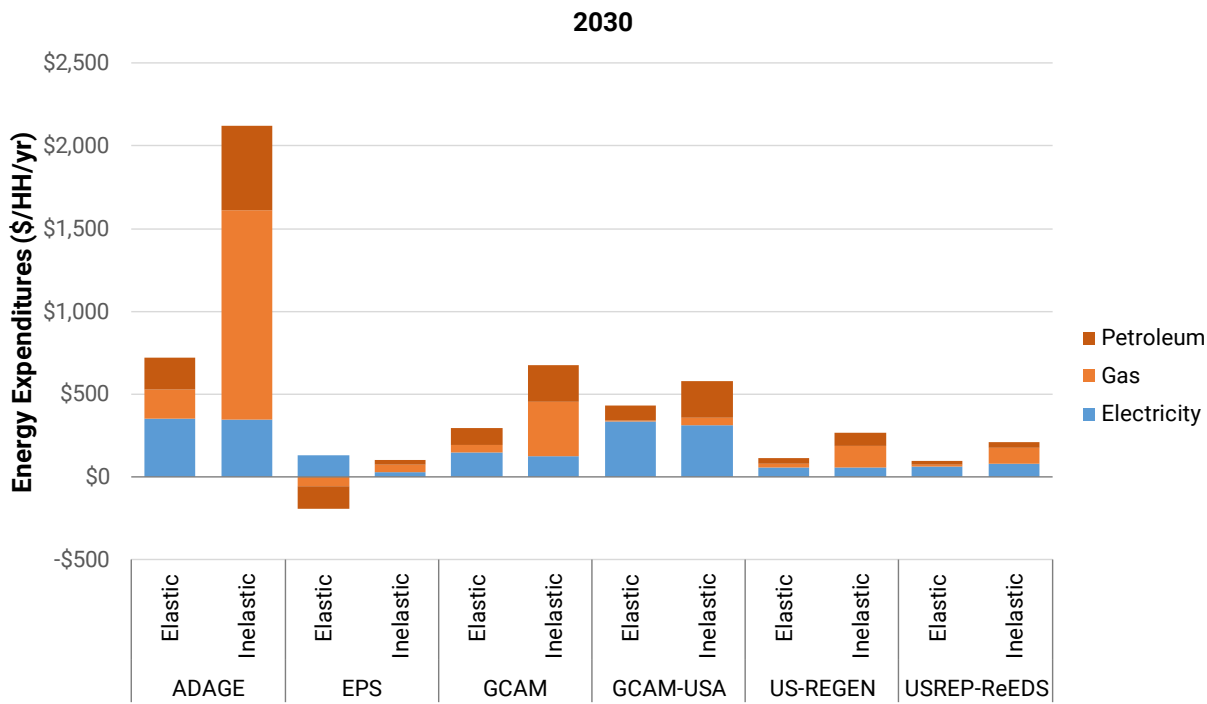

**Figure 10. Change in annual energy expenditures by fuel between the 0by50 and reference scenarios for lowest-income households in 2030.** Values are shown assuming elastic demand (from earlier results) and inelastic demand.

*3.3. Impacts for Models with Income Classes*

We can also directly examine outputs for the subset of models that report variables across income quintiles. With this approach, since the reported quantities of energy and overall goods consumed are endogenous, the outputs consider not only the impact of changes in energy prices on existing consumption patterns but also the impact of changes to consumption patterns in response to these new prices. We can also look at the overall welfare changes inclusive of all elements entering the household utility functions, including goods, services, and leisure. In this section, we discuss outputs from the WITCH model, which is the only participating model in this study to report outputs by income.[18]

---

[18] Though USREP-ReEDS reports variables by income class explicitly, we did not include it in the corresponding discussion in this section due to not reporting baseline income and consumption variables that would have allowed us to normalize for comparison across models.

The net-zero target is modeled as a carbon cap with a corresponding carbon price and revenue. All models in EMF 37 assume that the revenue is redistributed evenly on a per-capita basis. Moreover, the WITCH model also implements a scenario in which carbon revenues are redistributed in a distributionally neutral manner. That is, it is implicitly assumed that carbon revenues do not affect the distribution at all (hence labeled "neutral"), for instance, if revenues are used to reduce government debt levels. The presence of this additional revenue reuse assumption allows us to look at the impact of the net-zero target itself versus the rebating of the revenue from the modeled carbon price. However, by 2050, impacts are primarily the result of emissions reductions, as opposed to revenue redistribution. The income classes in WITCH are modeled in a coupled household optimization problem on intertemporal consumption allocation between energy (transportation and residential energy use), food, and other goods consumption, calibrated on survey data, with endogenous consumption, wages, savings, and wealth dynamics [30].

Figure 11 presents changes in energy consumption under a net-zero target relative to the no-policy case. As in Figure 8, lower-income households face the largest percentage increase in energy costs relative to their higher-income counterparts, with cost increases of approximately $200 annually per household for the lowest income quintile, amounting to roughly 1% of their baseline income. This suggests that the net-zero policy as modeled is regressive when looking at energy burden alone. Note that in the WITCH model, energy demand is captured at the macro level and does not interact with the distributional model, leading to energy consumption across income groups to be identical across revenue reuse assumptions.

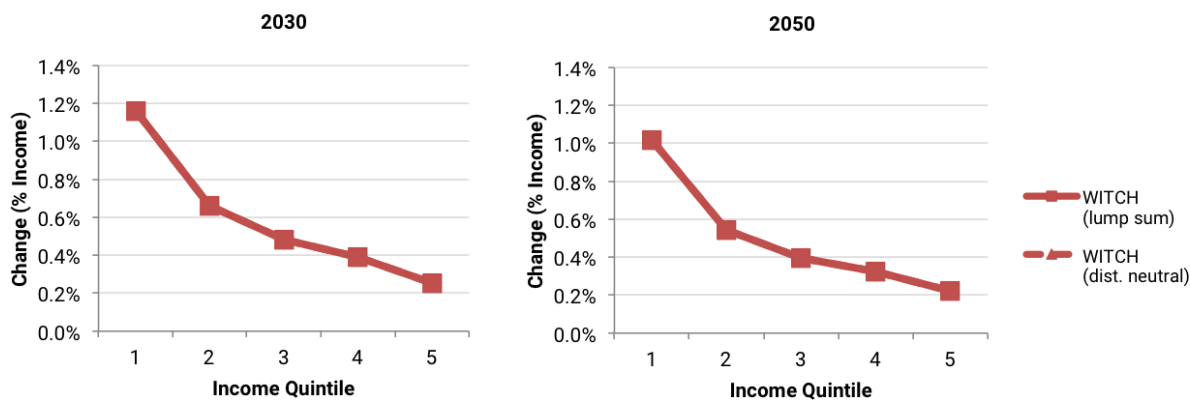

**Figure 11. Change in energy consumption under net-zero target relative to no-policy case across income quintiles by revenue reuse assumption (% of income).** "Lump sum" refers to carbon revenues being rebated lump-sum on a per-capita basis; revenues are reused in a manner that does not affect distribution under "dist. neutral." Results are shown for 2030 (left) and 2050 (right).

We can also look at how these models suggest that net-zero targets will impact overall levels of consumption on all goods and services between income groups, as represented in Figure 12. Looking first at the distributionally neutral revenue recycling assumption, we can see that the net-zero policy alone is slightly regressive and reduces consumption across all income groups by approximately 1% in 2030 (left panel) and 4% in 2050 (right panel). In 2030, adding a lump-sum per-capita rebate makes the policy highly progressive, with consumption for the lowest quintile increasing by as much as 10% due to the large amount of carbon revenues rebated relative to their baseline income. By 2050, the scenarios converge such that the policy is slightly regressive, owing to the limited carbon revenue once the economy reaches net-zero emissions.

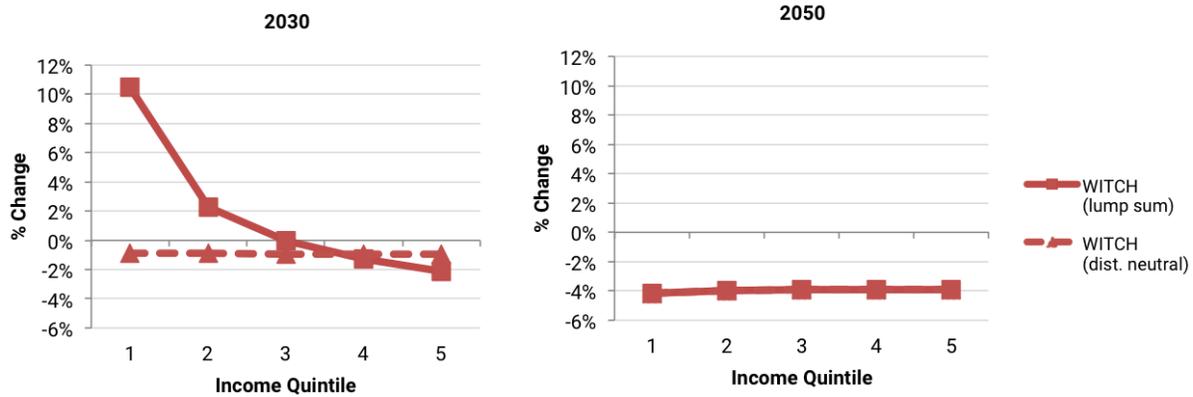

**Figure 12. Change in total consumption under net-zero target relative to no-policy case across income quintiles by revenue reuse assumption (% change).** Results are shown for 2030 (left) and 2050 (right). "Lump sum" refers to carbon revenues being rebated lump-sum on a per-capita basis; revenues are reused in a manner that doesn't affect distribution under "dist. Neutral." "Total consumption" includes all goods and services.

Finally, the models that explicitly report results by income quintile allow us to examine the distributional impact of the policy regarding total policy cost in terms of equivalent variation, incorporating not only the welfare impacts associated with changes in households' consumption of goods and services but also the utility associated with leisure consumption. Like total consumption, we can see in Figure 13 that the net-zero policy is welfare-reducing and slightly regressive absent the lump-sum transfer, but with the transfer becomes highly progressive in 2030. As above, the scenarios converge to be slightly regressive in 2050 with limited carbon revenues.

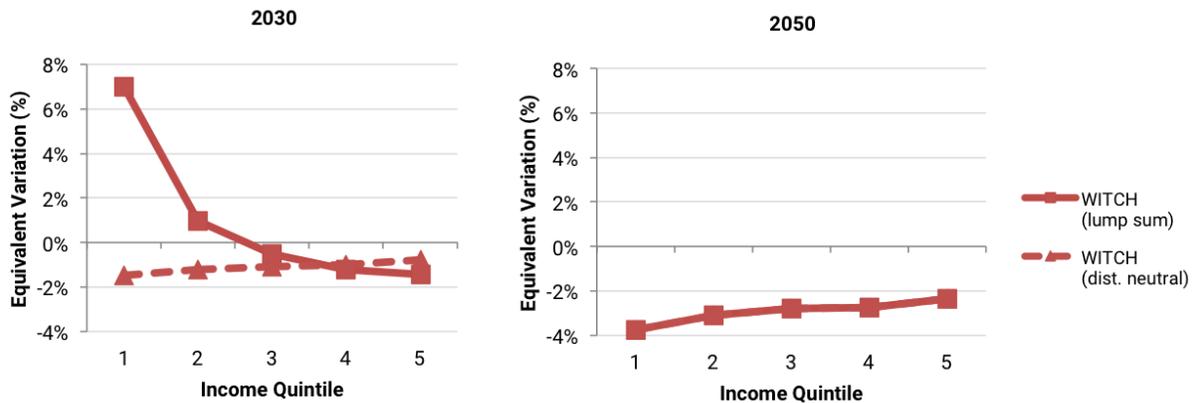

**Figure 13. Policy cost of net-zero target in equivalent variation relative to no-policy case (% of expenditures).** Results are shown for 2030 (left) and 2050 (right).

## 4. Discussion and Conclusions

This study uses a novel linking of detailed energy-economic models with consumer data to characterize the distributional effects of energy system transformations and proposed policies across income classes. The study uses seven models with harmonized scenarios reaching economy-wide, net-zero $CO_2$ emissions across the U.S. in 2050.

There are several key takeaways from the analysis:

- Climate policy design, mainly whether revenues are raised and how they are used, has first-order impacts on distributional outcomes. Before accounting for recycled $CO_2$ revenues, the cap-based net-zero policies studied here are regressive, in that they have the most significant percentage cost impact on lower-income households. In other words, if the net-zero policy—as implemented through a cap on $CO_2$ emissions in most participating models—does not facilitate compensating household transfers, it may have a regressive impact. Depending on their design, lump-sum transfers can mute these impacts, especially for the lowest-income households, and make net-zero policies progressive for many models and periods. Net-zero policies lead to net rebates for many consumers, models, and periods. One participating model (EPS) uses a combination of policies to reach the net-zero targets, including standards, instead of a $CO_2$ constraint and also finds progressive outcomes. Many models in this study assume that carbon pricing does a large share of the heavy lifting to reach economy-wide net-zero $CO_2$, but if other instruments play more significant roles than is assumed in these scenarios or if carbon pricing has more exclusions, then government revenues may be more limited, which likely would lower the progressivity of the policy, as these comparisons illustrate.

- Models generally agree that direct energy expenditures for households will likely decline over time with reference and net-zero policies, partly due to electrification; however, there is variation in the extent of electrification and changes between a reference and net-zero policy.

- We illustrate how using annual income deciles for distributional analysis instead of expenditure deciles can overstate the progressivity of emissions policies by overweighting revenue impacts on the lowest-income deciles. Shifting to an expenditure-based analysis can more accurately estimate distributional impacts and flattens the distribution across income classes.

These findings suggest several areas for future work. First, the importance of revenue recycling suggests that additional scenarios are warranted around alternate policy designs, including different revenue recycling assumptions (similar to the sensitivities conducted in EMF 32 [25]), net-zero emissions across all greenhouse gases, and alternate technology assumptions. These scenarios also should assess the implications of Inflation Reduction Act incentives across models, which can have large emissions and energy system impacts and consequently effects on distributional outcomes [31, 32]. Additional scenarios also would be valuable for looking at alternative policy portfolios to reach economy-wide net-zero emissions. Second, given how low-carbon transitions often lead to air quality improvements and human health benefits [33, 34], additional work should investigate the distributional impacts of these changes, including in a multi-model setting. Third, other sensitivities should be conducted with net-zero GHG emissions by 2050, which is more stringent than the net-zero $CO_2$ policy investigated here.

There are several opportunities to refine energy models to examine equity-related impacts:

- Adding explicit income-related structural classes to track welfare implications

- Providing regional reporting to understand locational variation in equity impacts, given possible regional heterogeneity in the existing technological stock, fuel prices, climate, and other factors[19]

- Adding heterogeneity to end-use decisions to understand opportunities and barriers to technology adoption across income classes, different building types, current technologies, and climate zones

- Adding representations of the economy and tax code (e.g., by linking energy system models to computable general equilibrium models) to examine alternate revenue recycling scenarios

- Coupling energy system models with more detailed tools to examine air quality and other localized impacts

Another area of future work is the investigation of horizontal equity dimensions of net-zero policies [35]. Although this analysis focused on vertical equity across income classes, there is considerable variation in welfare changes within income classes due to heterogeneity in household energy consumption, even while average effects are relatively limited. Figure 14 shows changes in net expenditures as a share of income for the GCAM-USA model in 2050. Instead of presenting the average of households within income classes (as Figure 8 shows), this figure shows the changes for individual consumer units from the CEX microdata. The comparison illustrates variation in policy impacts *within* income classes (in addition to changes *across* income, as earlier results in this paper investigated in detail), suggesting that horizontal equity is important for future work.

---

[19] Models do not report at a regional level for this analysis and have different levels of aggregation (Table 1), which makes such regional comparisons challenging.

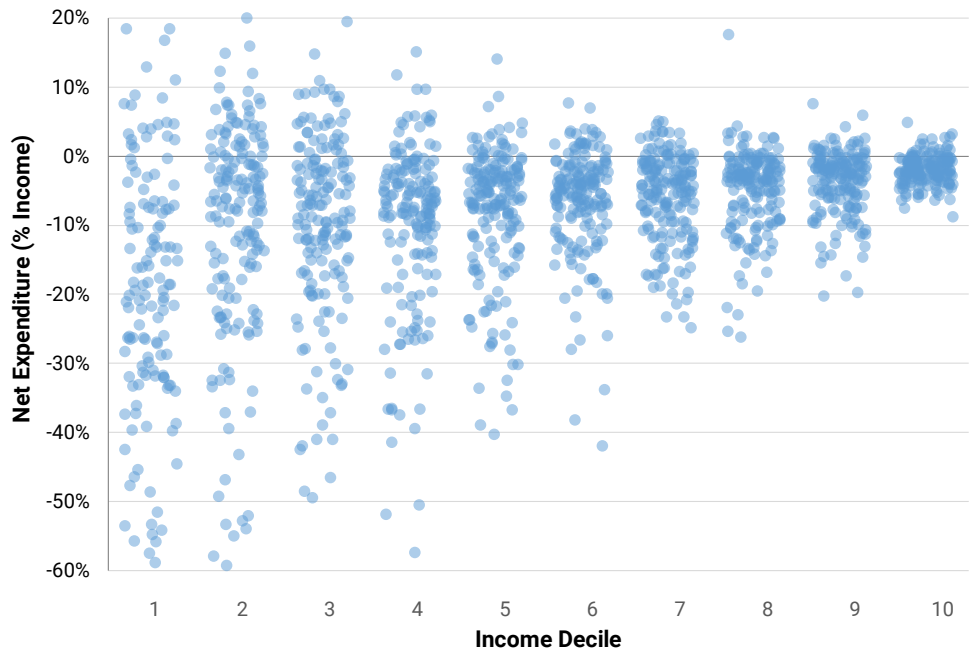

**Figure 14. Energy expenditures less recycled revenue as share of income under the net-zero scenario.** Results are shown for the GCAM-USA model in 2050. Points represent individual consumer units from the CEX microdata but do not show population weights.

**Acknowledgments**

The views expressed in this paper are those of the individual authors and do not necessarily reflect those of their respective institutions. J.E. and L.S. acknowledge funding from the European Union's Horizon 2020 research and innovation program under grant agreement number 101022622 (ECEMF) and from the European Research Council under grant agreement number 853487 (2D4D).

**Appendix**

Retail petroleum prices in Figure 15 illustrate differences across model assumptions in the reference scenario. Variation in prices under the 0by50 scenario, which include carbon prices, largely reflect differences in magnitudes of $CO_2$ prices (Figure 1).

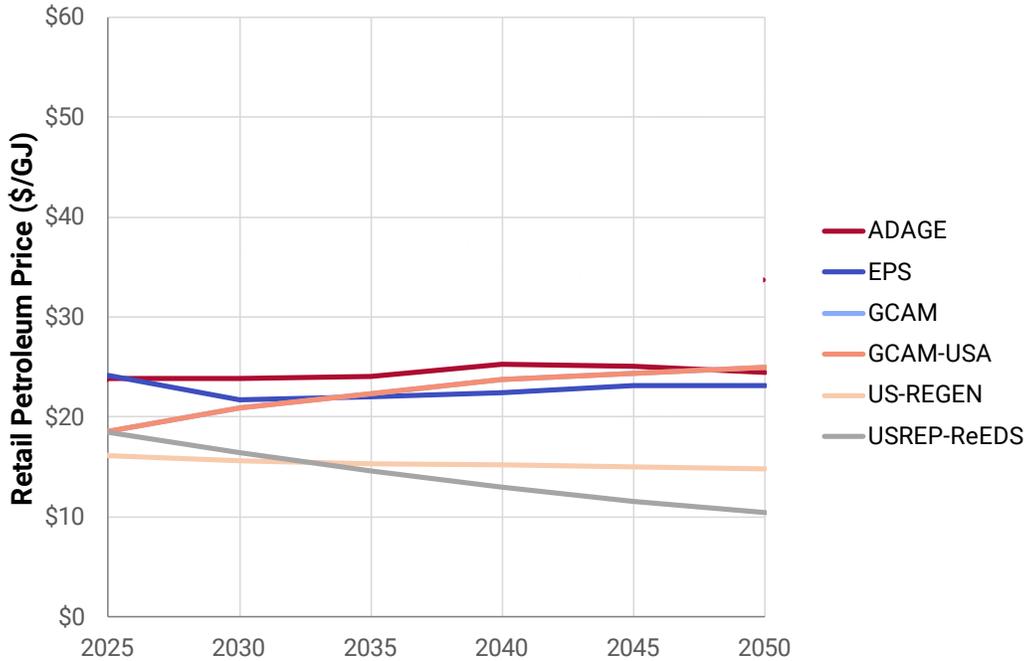

**Figure 15. Retail petroleum prices over time by model.** Prices are shown for the reference (solid lines) and 0by50 (dotted line) scenarios. GCAM and GCAM-USA have the same petroleum prices.

Trajectories of annual net revenues are shown in Figure 16.

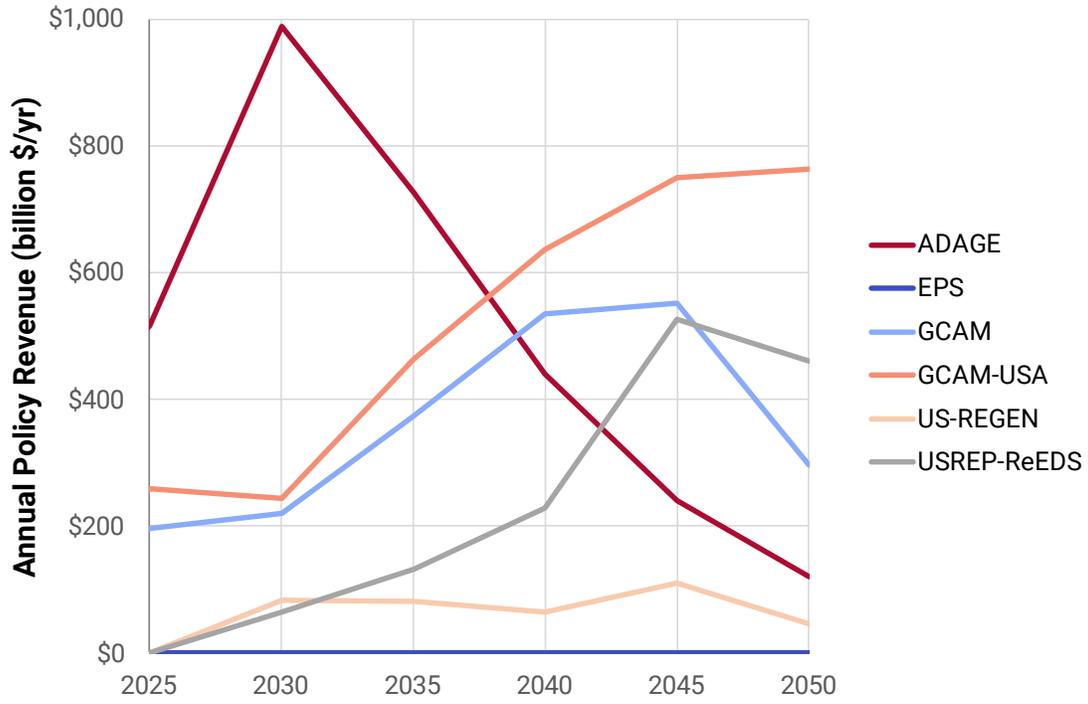

**Figure 16. Annual revenue from net-zero CO$_2$ policy by model over time.** Revenues are shown net of CDR payments.

30